\documentclass[12pt,a4paper]{article}
\usepackage{epsfig}
\usepackage{latexsym}

\newcommand{\Jbar}{\overline{J}}
\newcommand{\Jbarcl}{\widetilde{\Jbar}}
\newcommand{\Jcl}{\widetilde{J}}
\newcommand{\Nhat}{\hat{N}}

\def\Nhcdlow{\Nhat_{\ulcd}}
\def\Ncdlow{N_{\ulcd}}

\def\Nhcdup{\Nhat^{\ulcd}}

\def\Jbzero{\Jbar^{[\ulcd]}}
\def\Jzero{J^{[\ulcd]}}

\def\Jalh{J^{\alh}}
\def\Jal{J^\al}
\def\Jdl{J^\dl}
\def\Jdlh{J^{\h\dl}}

\def\JBal{\Jbar^{\al}}
\def\JBalh{\Jbar^{\alh}}

\def\JBdl{\Jbar^{\dl}}
\def\JBdlh{\Jbar^{\h\dl}}

\def\Ja{J^{\ula}}

\def\Jd{J^{\ul d}}
\def\JBa{\Jbar^{\ula}}
\def\JBb{\Jbar^{\ulb}}
\def\JBd{\Jbar^{\ul d}}

\newcommand{\ula}{\underline{a}}
\newcommand{\ulb}{\underline{b}}

\newcommand{\ulcd}{\underline{cd}}
\newcommand{\ulab}{\underline{ab}}

\newcommand{\h}{\hat}
\newcommand{\ul}{\underline}
\newcommand{\ga}{\{}
\newcommand{\gc}{\}}
\newcommand{\fraz}{\frac{1}{2}}
\def\pbar{\overline{\p}}
\def\p{\partial}
\def\lpbar{\overleftarrow{\pbar}}
\def\rpbar{\overrightarrow{\pbar}}
\def\lp{\overleftarrow{\p}}
\def\rp{\overrightarrow{\p}}
\def\nabar{\overline{\nabla}}
\def\yb{\overline{y}}
\def\zb{\overline{z}}
\def\omb{\overline{\om}}

\def\be{\begin{equation}}
\def\ee{\end{equation}}
\def\bea{\begin{eqnarray}}
\def\eea{\end{eqnarray}}

\def\al{\alpha}
\def\bt{\beta}
\def\bth{\hat{\beta}}
\def\alh{\hat{\alpha}}
\def\gm{\gamma}                \def\Gm{\Gamma}
\def\dl{\delta}                

\def\lm{\lambda}               \def\Lm{\Lambda}
               
\def\om{\omega}               
\def\na{\nabla}

\def\gmcd{\gm_{\ulcd}}

\def\gma{\gm_{\ula}}

\def\eab{\eta_{\al\bth}}
\def\ebai{\eta_{\bth\al}}
\def\eabi{\eta_{\alh\bt}}
\def\etaab{\eta_{\ulab}}

\def\Xal{X^\al}
\def\Xalh{X^{\alh}}
\def\Xdl{X^\dl}
\def\Xdlh{X^{\h\dl}}
\def\Xa{X^{\ula}}
\def\Xb{X^{\ulb}}
\def\Xd{X^{\ul d}}
\def\Xbt{X^{\bt}}
\def\Xbth{X^{\bth}}

\begin{document}
\bigskip\begin{titlepage}
\begin{flushright}
UUITP-11/06\\
\end{flushright}
\vspace{1cm}
\begin{center}
{\Large\bf Operator Product Expansion for Pure Spinor \\Superstring on $AdS_5\times S^5$\\}
\end{center}
\vspace{3mm}
\begin{center}
{\large
Valentina \ Giangreco M. Puletti{$^1$}} \\
\vspace{5mm}
Institutionen f\"or Teoretisk Fysik, \\Uppsala Universitet, Box 803, SE-751 08
Uppsala, Sweden \\
\vspace{5mm}
{\tt
{$^1$}valentina.giangreco@teorfys.uu.se\\
}
\end{center}
\vspace{5mm}
\begin{center}
{\large \bf Abstract}
\end{center}
\noindent The tree-level operator product expansion coefficients of
the matter currents are calculated  in the pure spinor formalism for
type IIB superstring in
the $AdS_5\times S^5$ background.\vfill
\begin{flushleft}
July 2006
\end{flushleft}
\end{titlepage}

%\maketitle

%\abstract{The tree-level operator product expansion coefficients of
%the matter currents are calculated  in the pure spinor formalism for
%type IIB superstring in
%the $AdS_5\times S^5$ background. %with Ramond-Ramond flux
%}

%\keywords{string theory, pure spinors.}

\newpage
\section{Introduction and Summary}
Type IIB superstring in an $AdS_5\times S^5$ background is
conjectured to be dual to $\mathcal{N}=4$ super Yang-Mills theory
in $D=4$ dimensions. To fully exploit the duality one would need
to solve the world-sheet sigma model with the  $AdS_5\times S^5$
target-space at the quantum level, and in particular to understand
its operator algebra. As a first step toward this ambitious goal,
we analyze in this paper the tree-level OPE of matter currents on
the world-sheet.

The type IIB superstring in the $AdS$ space with Ramond-Ramond flux
can be formulated as a sigma model with the target space which is
the supermanifold $\frac{PSU(2,2|4)}{SO(4,1)\times SO(5)}$. The
action in the Green-Schwarz (GS) formalism is known~\cite{tseytlin}.
The pure spinor (PS) version was proposed in~\cite{B0001035}. In
both these approaches the target-space supersymmetry is manifest.
However in the GS formulation where the world-sheet action is
classically $\kappa$-invariant, the covariant quantization of the
sigma model is rather complicated due to non-linearities and because
gauge-fixing the $\kappa$-symmetry leads to fermionic second-class
constraints.  In the PS formalism  proposed by Berkovits, the
manifest target-space supersymmetry of the world-sheet sigma model
is obtained using essentially free fields. In this approach the main
ingredients are the commuting left and right-moving space-time
spinors $\lm^\al$ and $\h\lm^{\alh}$, which play the role of ghost
variables and satisfy the \emph{pure spinor} constraint:
\bea \label{pscondition} \lm\gm^{\ula} \lm=\h\lm\gm^{\ula}\h\lm=0 .
\eea The world-sheet superstring action is classically
BRST-invariant in the Berkovits formulation: due to the presence of
a kinetic term for the fermionic currents the $\kappa$-symmetry of
the GS superstring action is spoiled and replaced by a BRST-like
symmetry whose charges are constructed from fermionic constraints
and pure spinors. In both these formalisms an infinite set of
non-local classically conserved charges has been found, which highly
suggest the integrability of the model~\cite{bena,V0307018}. At the
classical level these non-local charges have been shown to be
$\kappa$-invariant in the GS formalism and BRST-invariant in the PS
formalism~\cite{B0409159}.
%The approach proposed by Berkovits is really promising.
 By cohomology arguments it was proved that the BRST invariance
 of the PS superstring action survive at the quantum level~\cite{B0411170}.
Moreover the superstring action was explicitly proved to be one-loop
conformally invariant in~\cite{V0210064}. The  (classical) current
algebra in the hamiltonian formalism was analyzed in~\cite{bianchi}.
Here we will compute the operator product expansion (OPE) of the
matter currents.

In section 2 the effective action for the fluctuations fields is
derived. The results for the OPEs is presented in section 3. Our
notation is summarized in the appendices, where we also give some
calculational details.

\section{The action}
In the pure spinor formalism the sigma model action describing an
$AdS_5\times S_5$ background with Ramond-Ramond flux
is~\cite{B0001035,V0210064,B0009168,V0307018,B0411170}
\bea
\label{initialaction} S_{AdS}&=& \frac{1}{\al^2}\int d^2 z
\ga <J_2,\Jbar_2>+\frac{3}{2}<J_3,\Jbar_1>+\fraz <\Jbar_3,J_1>\gc
+\cr &+& \frac{1}{\al^2}\int d^2z \ga
\Ncdlow\Jbzero+\Nhcdlow\Jzero+\fraz \Ncdlow \Nhcdup\gc
+\frac{1}{\al^2}(S_\lm +S_{\h\lm}),
\eea
where $<,>$ is the bilinear form expressed in terms of the super-trace, $J=J^A T_A$, $\Jbar=\Jbar^A T_A$, with $T_A$ the generators of the super-algebra and $J^A=(g^{-1}\p g)^A$ and $\Jbar^A=(g^{-1}\pbar g)^A$ are the left
invariant (super) currents constructed from $g(x,\theta,\h\theta)$ which are
elements of the super-coset $\frac{PSU(2,2|4)}{ SO(4,1)\times SO(5)}$ and $(x,\theta,\h\theta)$ parameterize the $D=10$, $N=2$ superspace.\\
$\Ncdlow=\fraz\om\gmcd\lm$ and $\Nhcdlow=\fraz \h\om\gmcd\h\lm$
are the $SO(4,1)\times SO(5)$ components of the Lorentz currents
for the pure spinor ghosts $\lm^\al$ and $\h\lm^{\alh}$ and their
conjugate momenta $\om_\al$ and $\h\om_{\alh}$, respectively.
$S_\lm$ and $S_{\h\lm}$ are the free field actions for the pure
spinors in the flat background.\\
The action is manifestly invariant under global $PSU(2,2|4)$
transformations which act by left multiplication on the coset
supergroup elements and it is invariant under local $SO(4,1)\times
SO(5)$ gauge transformations which act by right multiplication on
$g$.\footnote{Under a local gauge $SO(4,1)\times SO(5)$
transformation with parameter $\xi\in \mathcal{H}_0$, $J_i$ and
$\Jbar_i$ transform as $\dl J_i=[J_i,\xi]$, $\dl
\Jbar_i=[\Jbar_i,\xi]$, while $J_0$ and $\Jbar_0$ transform as a
connection $\dl J_0=\p\xi+[J_0,\xi]$, $\dl
\Jbar_0=\pbar\xi+[\Jbar_0,\xi]$, the ghost currents as $\dl
N=[N,\xi]$, $\dl \Nhat=[\Nhat,\xi]$, and the pure spinors and
their conjugate momenta transform as $\dl \lm=[\lm,\xi]$, $\dl
\om=[\om,\xi]$, analogously for the hatted
spinors,~\cite{V0307018,B0411170}.}
The coupling constant is $\al=(\lm)^{-\frac{1}{4}}=(Ng_s)^{-\frac{1}{4}}$ and the
coefficients of the action are fixed in such a way that the action
is gauge invariant under local $SO(4,1)\times SO(5)$ transformations,
according to the metric and the structure constant normalized as
in the appendix.

Using the background field method~\cite{B9907200,V0210064,PSL},
one can compute the one-loop effective action. The mapping $g$ is
parameterized as a classical background plus quantum fluctuations
around the background: $g=\tilde{g}e^{\al X}$. The gauge
invariance of the original action can be used to fix $X\in \mathcal{G/H}_0$. Plugging $g=\tilde{g}e^{\al X}$ in
$J$, $\Jbar$ and expanding up to the $\al^2$ order, the matter
currents are given in terms of the $X$ fields by:
\bea \label{expansionj} &&J_i=\Jcl_i +\al \big( \p X_i+[\Jcl,X]_i
\big)+\frac{\al^2}{2}\big([[\Jcl,X],X]_i+[\p X,X]_i\big)+...\cr
&&\Jbar_i=\Jbarcl_i +\al \big( \pbar X_i+[\Jbarcl,X]_i
\big)+\frac{\al^2}{2}\big([[\Jbarcl,X],X]_i+[\pbar
X,X]_i\big)+..., \eea
where $i=1,2,3$ labels the elements of the subalgebras
$\mathcal{H}_i$, i.e. $J_i\equiv J_{|\mathcal{H}_i}$, $\Jcl$ and $\Jbarcl$ are the classical currents,
$\Jcl=\tilde{g}^{-1}\p\tilde{g}$,
$\Jbarcl=\tilde{g}\pbar\tilde{g}$. \\
Though $X \in \mathcal{G/H}_0$, the fluctuations can contain the
gauge fields since the commutators in (\ref{expansionj}) can
contain $J_0$ and $\Jbar_0$. Thanks to the gauge invariance, the
effective action is independent of them~\cite{B9907200}, thus they
can be gauged away, i.e. $[J_0,X_i]=[\Jbar_0,X_i]=0$ for any
$X_i$.
Furthermore $J_0$ and $\Jbar_0$ can have quantum fluctuations
according to:
\bea \label{expansionj0} &&J_0=\tilde{J}_0 +\al [\tilde{J}, X]_0
+\frac{\al^2}{2}\big( [\p X,X]_0+ [[\tilde{J},
X],X]_0\big)+...,\cr &&\Jbar_0=\tilde{\Jbar}_0 +\al
[\tilde{\Jbar}, X]_0 +\frac{\al^2}{2}\big( [\pbar X,X]_0+
[[\tilde{\Jbar}, X],X]_0\big)+.... \eea We will treat the Lorentz
ghost currents as external ones.

Since we want to know the tree-level OPE for the matter currents,
we need to compute the action for the $X$ fluctuations only to the
first order in the external currents. Plugging the expansion of
the currents (\ref{expansionj}) and (\ref{expansionj0}) in the
action (\ref{initialaction}), one gets terms of zeroth order in
the $X$ fields, which are the action for the background fields,
linear terms in $X$, which vanish by general arguments of QFT, and
second order terms in the fluctuations. Thus keeping all the terms
of $\al^2$ order and neglecting all the contributions which are of
the second order in the classical currents\footnote{For a
dimensional analysis this implies neglecting also the terms of
order $~\p J$.}, one obtains the following action:
\bea
\label{myaction} S&=&  \int d^2z \big\ga -\etaab \Xa\p\pbar
\Xb-\eab\Xal\p\pbar\Xbth-\ebai\Xbth\p\pbar\Xal\big\gc+\cr &+&\
\int d^2z \big\ga \Xal\big[\fraz \etaab
f^{\ulb}_{\al\bt}\big(\lpbar\Ja +\Ja \rpbar\big)\big]\Xbt+
\Xalh\big[\fraz \etaab
f^{\ula}_{\alh\bth}\big(\lp\JBb+\JBb\rp\big)\big]\Xbth+\cr &+&
\Xa
\big[\eab
f^{\bth}_{\ula\bt}\big(\lpbar\Jal+\Jal\rpbar\big)\big]\Xbt
+\Xa\big[\eabi
f^{\bt}_{\ula\bth}\big(\lp\Jbar^{\alh}+\Jbar^{\alh}\rp\big)\big]\Xbth+\cr
&+&\Xa\big[-\frac{1}{4}f^{[\ulcd]}_{\ulab}\big(\lpbar
\Ncdlow+\Ncdlow\rpbar\big)-\frac{1}{4}f^{[\ulcd]}_{\ulab}\big(\lp\Nhcdlow+\Nhcdlow\rp\big)\big]\Xb+\cr
&+& \Xal\big[-\fraz
f^{\ulcd}_{\al\bth}\big(\lpbar\Ncdlow+\Ncdlow\rpbar\big)-\fraz
f^{\ulcd}_{\al\bth}\big(\pbar\Nhcdlow+\Nhcdlow\pbar\big)\big]\Xbth\big\gc
\eea
All the currents that are present in the action (\ref{myaction})
are the classical ones (the $\tilde{}$ is omitted in the notation
on what follows). From a diagrammatic point of view this means
considering all the tree-level diagrams, i.e. with one insertion
of the external current, either matter current $J^A$, $\Jbar^A$ or
Lorentz ghost current $N_{\ulcd}$, $\Nhat_{\ulcd}$.
\section{The tree-level interactions}
In order to read off the propagators for the quantum fluctuations
one has to invert perturbatively the operator between the X's. The
kinetic term is given by the (super) matrix:
\begin{eqnarray}
\label{Amatrix}
A=
\left[\begin{array}{ccc}
\etaab(-\p\pbar) & 0 & 0 \\
0 & 0 & \eab(-\p\pbar) \\
0 & \eabi(-\p\pbar) & 0\\
\end{array} \right] .
\end{eqnarray}
The inverse of this matrix is given by:
\bea
\label{Ainverse}
A^{-1}=
\left[\begin{array}{ccc}
\eta^{ab}(-\p\pbar)^{-1} &0&0 \\
0 & 0& \eta^{\bt\alh}(-\p\pbar)^{-1}\\
0 & \eta^{\bth \al}(-\p\pbar)^{-1} & 0\\
\end{array}\right].
\eea where $(-\p\pbar)^{-1}$ is  formally the "free" propagators,
namely is given by: $\label{freepropagator} (-\p\pbar)^{-1}=
-\frac{1}{2\pi}\log|y-z|^2$. The coefficients in front of the
propagator is fixed by the differential equation
$\p\pbar\log|z|^2=2\pi\dl^{(2)}(z,\zb)$\footnote{The $\dl$
function in the complex plane is normalized as
in~\cite{Polchinski}.}. In this way the integral of the $\dl$
function is normalized to 1, $\int d^2 z \dl^{(2)}(z,\zb)=1$.
%This
%normalization is fixed by using the classical equations of motion
%(\ref{classicaleom}).

If the operator can be represented as a sum
of two matrices $A$ and $V$, the inverse of the matrix is
perturbatively:
\bea
(A+V)^{-1}=A^{-1}-(A^{-1}VA^{-1})+...
\eea
where V is the matrix containing the tree-level interaction with
the matter and the Lorentz ghost currents. The entries of the matrix V are
just the terms containing the currents in the action
(\ref{myaction}), the terms off-diagonal divided by 1/2:
\bea \label{Vmatrix}
&&V_{11}=
-\frac{1}{4}(f^{[\ulcd]}_{\ulab}(\lpbar \Ncdlow+\Ncdlow\rpbar)+
f^{[\ulcd]}_{\ulab}(\lp\Nhcdlow+\Nhcdlow\rp)) \cr
&&V_{12}= \fraz
\eta_{\rho\hat{\rho}} f^{\h\rho}_{\ula\bt}(\lpbar
J^{\rho}+J^{\rho}\rpbar) \cr
&&V_{13}= -\fraz \eta_{\rho\hat{\rho}} f^{\rho}_{\ula\bth}(\lp
\Jbar^{\h\rho}+\Jbar^{\h\rho} \rp)\cr
&&V_{22}=\fraz \eta_{\ulab}
f^{\ulb}_{\al\bt}(\lpbar J^{\ula}+J^{\ula}\rpbar)\cr
&&V_{23}=
-\frac{1}{4}(f^{\ulcd}_{\al\bth}(\lpbar N_{\ulcd}+
N_{\ulcd}\rpbar)+f^{\ulcd}_{\al\bth}(\lp \Nhat_{\ulcd}+
\Nhat_{\ulcd}\rp))\cr
&&V_{33}=\fraz\eta_{\ulab}f^{\ula}_{\alh\bth}(\lp
\Jbar^{\ulb}+\Jbar^{\ulb}\rp). \eea

\section{OPE}
The general expression for the OPE of the currents is at the order
considered:
\bea \label{opej}
J^A(y)J^B(z) &=&
<\widetilde{J}^A(y)\widetilde{J}^B(z)>+\al^2 \Big( <\p X^A(y) \p
X^B(z)>+\cr &&+ <\p
X^A(y)[\widetilde{J},X]^B(z)>+<[\widetilde{J},X]^A(y) \p X^B (z)>+
\mathcal{O}(J^2) \Big).\cr && \eea
The currents are taken normal-ordered to avoid the contractions on
the same points. The classical terms given by the propagator of
two currents in (\ref{opej}) will be not considered and the
contractions on the last two terms in (\ref{opej}) must be done
keeping in mind that these contributions are already at the
tree-level order, namely they already contain an external leg. On
what follows all the currents that appear in the l.h.s. of the OPE
are the classical ones. The results are proportional to the
coupling constant $\al^2$, this overall factor is omitted in the
final result as well as the classical term.
%
%LISTA DEGLI OPE DELLE CORRENTI
%
%
%\subsection{$J_2 J_2$}

%In the case of two bosonic matter current the contribution comes
%only from $<\p X_2\p X_2>$ in (\ref{opej}), since the commutators in $\mathcal{H}_2$ are
%$[J_1,X_1]$ or $[J_3,X_3]$ and in (\ref{Ainverse}) we cannot
%contract $X_2$ with $X_3$ or $X_1$\footnote{In the appendix all
%the OPE's are computed explicitly.}.
%
\bea
\Ja(y)\Jd(0)&\simeq & (cl.)+ \al^2 ( <\p \Xa (y)\p
\Xd(0)>+...)\simeq \cr &\simeq &
-\frac{\eta^{\ul{ad}}}{2\pi}\frac{1}{y^2}+\frac{1}{4\pi}\eta^{[\ul{ad}][\ul{ef}]}\Nhat_{\ul{ef}}(0)\frac{\yb}{y^2}-\frac{1}{4\pi
y}\eta^{[\ul{ad}][\ul{ef}]} N_{\ul{ef}}(0).\cr && \eea
%
%
% J_2 Jbar_2
%
\bea \label{j2jbar2}
\Ja(y)\JBd(0) &\simeq &(cl) +\al^2 ( <\p \Xa
(y)\pbar \Xd(0)>+...)\simeq\cr &\simeq &
-\frac{1}{4\pi\yb}\eta^{[\ul{ad}][\ul{ef}]}N_{\ul{ef}}(0)-\frac{1}{4\pi
y}\eta^{[\ul{ad}][\ul{ef}]}\Nhat_{\ul{ef}}(0). \cr &&\eea
%
%
% Jbar_2 J_2
%
\bea \label{jbar2j2} \JBa(y)\Jd(0) &\simeq &(cl) +\al^2 ( <\pbar
\Xa (y)\p \Xd(0)>+...)\simeq \cr &\simeq &
-\frac{1}{4\pi\yb}\eta^{[\ul{ad}][\ul{ef}]}N_{\ul{ef}}(0)-\frac{1}{4\pi
y}\eta^{[\ul{ad}][\ul{ef}]}\Nhat_{\ul{ef}}(0).\cr && \eea
%
%
% Jbar_2 Jbar_2
%
\bea \label{jbar2jbar2} \JBa(y)\JBd(0)&  \simeq & (cl) +\al^2 (
<\pbar \Xa (y)\p \Xd(0)>+...)\simeq \cr &\simeq &
-\frac{1}{2\pi\yb^2}\eta^{\ul{ad}}+\frac{1}{4\pi}\frac{y}{\yb^2}\eta^{[\ul{ad}][\ul{ef}]}
N_{\ul{ef}}(0) -\frac{1}{4\pi \yb}
\eta^{[\ul{ad}][\ul{ef}]}\Nhat_{\ul{ef}}(0). \eea

%\subsection{$J_2 J_1$}

%In the case of the OPE between currents with bosonic and fermionic
%indices we have also the contribution from the last two terms in
%(\ref{opej}).

\bea \Ja(y)\Jdl(0)&\simeq & \al^2 (<\p\Xa(y)\p\Xdl(0)>+
<\p\Xa(y)[J_3,X_2]^\dl (0)>+
<[J_3,X_3]^{\ula}(y)\p\Xdl(0)>+...)\simeq \cr & \simeq &
\frac{1}{2\pi}\frac{\yb}{y^2}f^{\ula}_{\h\gm\h\rho}\eta^{\h\gm\dl}\Jbar^{\h\rho}(0)+
\frac{1}{\pi y}
f^{\ula}_{\h\rho\h\gm}\eta^{\h\gm\dl}J^{\h\rho}(0). \eea
%
%
%For the other components of the OPE the procedure is the same and the results are the
%following:
%
% J_2 Jbar_1
%
\bea \label{j2jbar1} \Ja(y)\JBdl(0)  &\simeq & \al^2
(<\p\Xa(y)\pbar\Xdl(0)>+ <\p\Xa(y)[\Jbar_3,X_2]^\dl (0)>+
<[J_3,X_3]^{\ula}(y)\pbar\Xdl(0)>+...)\simeq \cr  &\simeq &
\frac{1}{2\pi
\yb}f^{\ula}_{\h\gm\h\rho}\eta^{\h\gm\dl}J^{\h\rho}(0). \cr &&
\eea
%
%
% Jbar_2 J_1
%
\bea \label{jbar2j1}
\JBa(y)\Jdl(0) &\simeq &  \al^2
(<\pbar\Xa(y)\p\Xdl(0)>+<\pbar\Xa(y)[J_3,X_2]^\dl
(0)>+<[\Jbar_3,X_3]^{\ula}(y)\p\Xdl(0)>+...)\simeq\cr
&\simeq & \frac{1}{2\pi
\yb}f^{\ula}_{\h\gm\h\rho}\eta^{\h\gm\dl}J^{\h\rho}(0). \eea
%
%
% Jbar_2 Jbar_1
%
\bea \label{jbar2jbar1} \JBa(y)\JBdl(0)&\simeq &\al^2
(<\pbar\Xa(y)\pbar\Xdl(0)>+ <\pbar\Xa(y)[\Jbar_3,X_2]^\dl
(0)>+<[\Jbar_3,X_3]^{\ula}(y)\pbar\Xdl(0)>+...)\simeq\cr
&\simeq &
\frac{1}{2\pi
\yb}f^{\ula}_{\h\gm\h\rho}\eta^{\h\gm\dl}\Jbar^{\h\rho}(0).
\eea
%

%\subsection{$J_2J_3$}

%Also in the OPE between the bosonic and the second fermionic
%current, the same kind of terms will contribute.
%
\bea
\Ja(y)\Jdlh (0)&\simeq & \al^2( <\p\Xa(y)\p\Xdlh
(0)>+<\p\Xa(y)[J_1,X_2]^{\h\dl}(0)>+<[J_1,X_1]^{\ula}(y)\p\Xdlh
(0)>+...) \simeq \cr &
 \simeq & \frac{1}{2\pi
y}f^{\ula}_{\rho\gm}\eta^{\gm\h\dl}J^\rho(0). \eea
%
%
%
% J_2 Jbar_3
%
\bea
\Ja(y)\JBdlh(0)&\simeq &\al^2 (<\p\Xa(y)\pbar\Xdlh(0)> +
<\p\Xa(y)[\Jbar_1,X_2]^{\h\dl}(0)>+<[J_1,X_1]^{\ula}(y)\pbar\Xdlh(0)>+...)\simeq \cr
&\simeq &
\frac{1}{2\pi y} f^{\ula}_{\rho\gm}\eta^{\gm\h\dl}\Jbar^\rho(0).
\eea
%
% Jbar_2 J_3
%
\bea
\JBa(y)\Jdlh(0)&\simeq & \al^2 (<\pbar\Xa(y)\p\Xdlh(0)> +
<\pbar\Xa(y)[J_1,X_2]^{\h\dl}(0)>+<[\Jbar_1,X_1]^{\ula}(y)\p\Xdlh(0)>+...)\simeq \cr
&\simeq &
\frac{1}{2\pi y} f^{\ula}_{\rho\gm}\eta^{\gm\h\dl}\Jbar^\rho(0).
\eea
%
%
% Jbar_2 Jbar_3
%
\bea
\JBa(y)\JBdlh(0)&\simeq &
\al^2 (<\pbar\Xa(y)\pbar\Xdlh(0)>+<\pbar\Xa(y)[\Jbar_1,X_2]^{\h\dl}(0)>+<[\Jbar_1,X_1]^{\ula}(y)\pbar\Xdlh(0)>+...)\simeq \cr
&\simeq &
\frac{1}{2\pi}\frac{y}{\yb^2}f^{\ula}_{\rho\gm}\eta^{\gm\h\dl}J^\rho(0)+
\frac{1}{\pi\yb} f^{\ula}_{\rho\gm}\eta^{\gm\h\dl}\Jbar^\rho(0).
\eea
%
%
%
%
%
%
%\subsection{$J_1 J_1$}
%
%This OPE involves two spinorial indices of the same type; again it
%receives contributions from the commutators present in
%(\ref{opej}), since $[J,X]_1$ can be $[J_3,X_2]$ or $[J_2,X_3]$ according to our gauge choice, and in the
%second case $X_3$ can be contracted with $X_1$ without external insertions.
%
\bea
\Jal(y)\Jdl(0)&\simeq &
\al^2(<\p\Xal(y)\p\Xdl(0)>+<\p\Xal(y)[J_2,X_3]^\dl(0)>+<[J_2,X_3]^\al(y)\p\Xdl(0)>+...)\simeq
\cr
&\simeq &
\frac{1}{2\pi}\frac{\yb}{y^2}f^{\al}_{\ul
l\h\gm}\eta^{\h\gm\dl}\Jbar^{\ul l}(0)+\frac{1}{\pi y}
f^{\al}_{\ul l\h\gm}\eta^{\h\gm\dl}J^{\ul l}(0).
\eea
%
%
% Jbar_1 J_1
%
\bea
\JBal(y)\Jdl(0)&\simeq &\al^2\ga
<\pbar\Xal(y)\p\Xdl(0)>+<\pbar\Xal(y)[J_2,X_3]^\dl(0)>+<[\Jbar_2,X_3]^\al(y)\p\Xdl(0)>+...\gc\simeq
\cr
&\simeq &  \frac{1}{2\pi\yb} f^{\al}_{\ul
l\h\gm}\eta^{\h\gm\dl}J^{\ul l}(0) .
\eea
%
%
% J_1 Jbar_1
%
\bea
\Jal(y)\JBdl(0) &\simeq &
\al^2(<\p\Xal(y)\pbar\Xdl(0)>+<\p\Xal(y)[\Jbar_2,X_3]^\dl(0)>+<[J_2,X_3]^\al(y)\pbar\Xdl(0)>+...)\simeq \cr
&\simeq &
\frac{1}{2\pi\yb}
f^{\al}_{\ul l\h\gm}\eta^{\h\gm\dl}J^{\ul l}(0).
\eea
%
%
% Jbar_1 Jbar_1
%
\bea \JBal(y)\JBdl(0) &\simeq &
\al^2(<\pbar\Xal(y)\pbar\Xdl(z)>+<\pbar\Xal(y)[\Jbar_2,X_3]^\dl(0)>+<[\Jbar_2,X_3]^\al(y)\pbar\Xdl(0)>+...)\simeq
\cr
&\simeq & \frac{1}{2\pi\yb} f^{\al}_{\ul
l\h\gm}\eta^{\h\gm\dl}\Jbar^{\ul l}(0).
\eea
%
%
%
%
%\subsection{$J_3 J_3$}
%
%
% J_3 J_3
%
\bea
\Jalh(y)\Jdlh(0)
&\simeq &\al^2 \ga
<\p\Xalh(y)\p\Xdlh(0)>+<\p\Xalh(y)[J_2,X_1]^{\h\dl}(0)>+<[J_2,X_1]^{\alh}(y)\p\Xdlh(0)>+...\gc\simeq
\cr
&\simeq &
\frac{1}{2\pi y} f^{\alh}_{\ul
l\gm}\eta^{\gm\h\dl}J^{\ul l}(0).
\eea
%
% Jbar_3 J_3
%
\bea
\JBalh(y)\Jdlh(0)  &\simeq &
\al^2 \ga
<\pbar\Xalh(y)\p\Xdlh(0)>+<\pbar\Xalh(y)[J_2,X_1]^{\h\dl}(0)>+<[\Jbar_2,X_1]^{\alh}(y)\p\Xdlh(0)>+...\gc\simeq \cr
&\simeq &   \frac{1}{2\pi y}f^{\alh}_{\ul
l\gm}\eta^{\gm\h\dl}\Jbar^{\ul l}(0).
\eea
%
% J_3 Jbar_3
%
\bea
\Jalh(y)\JBdlh(0) &\simeq &\al^2 \ga
<\p\Xalh(y)\pbar\Xdlh(0)> +
<\p\Xalh(y)[\Jbar_2,X_1]^{\h\dl}(0)>+<[J_2,X_1]^{\alh}(y)\pbar\Xdlh(0)>+...\gc \simeq \cr
&\simeq &
\frac{1}{2\pi y}f^{\alh}_{\ul l\gm}\eta^{\gm\h\dl}\Jbar^{\ul
l}(0).
\eea
%
% Jbar_3 Jbar_3
%
\bea
\JBalh(y)\JBdlh(0) &\simeq &\al^2 \ga
<\pbar\Xalh(y)\pbar\Xdlh(0)> +
<\pbar\Xalh(y)[\Jbar_2,X_1]^{\h\dl}(0)>+<[\Jbar_2,X_1]^{\alh}(y)\pbar\Xdlh(0)>+...\gc \simeq \cr
&\simeq &
\frac{1}{2\pi}\frac{y}{\yb^2}f^{\alh}_{\ul
l\gm}\eta^{\gm\h\dl}J^{\ul l}(0)+ \frac{1}{\pi\yb} f^{\alh}_{\ul
l\gm}\eta^{\gm\h\dl}\Jbar^{\ul l}(0).
\eea
%
%
%
%
%
%\subsection{$J_3 J_1$}
%
%
%Analogously to the case $J_2J_2$ OPE:
%
% J_3 J_1
%
\bea
\Jalh(y)\Jdl(0)& \simeq & (\textsl{cl}.) + \al^2
<\p\Xalh(y)\p\Xdl(0)>+... \simeq
\cr &\simeq & -\frac{1}{2\pi
y^2}\eta^{\alh\dl}+
\frac{1}{4\pi}\frac{\yb}{y^2}\eta^{\alh\bt}f^{[\ul{ef}]}_{\bt\h\gm}\eta^{\h\gm\dl}
\Nhat_{\ul{ef}}(0)- \frac{1}{4\pi
y}\eta^{\alh\bt}f^{[\ul{ef}]}_{\bt\h\gm}\eta^{\h\gm\dl}N_{\ul{ef}}(0).\cr &&
\eea
%
%
% Jbar_3 J_1
%
\bea
\JBalh(y)\Jdl(0) &\simeq & (\emph{cl.}) + \al^2 <\pbar\Xalh(y)\p\Xdl(0)>+... \simeq \cr
&\simeq & -\frac{1}{4\pi \yb}\eta^{\alh\bt}f^{[\ul{ef}]}_{\bt\h\gm}\eta^{\h\gm\dl}  N_{\ul{ef}}(0)-\frac{1}{4\pi y}\eta^{\alh\bt}f^{[\ul{ef}]}_{\bt\h\gm}\eta^{\h\gm\dl}\Nhat_{\ul{ef}}(0).
\eea
%
%
% J_3 Jbar_1
%
\bea
\Jalh(y)\JBdl(0)&\simeq &(\emph{cl.}) + \al^2
<\p\Xalh(y)\pbar\Xdl(0)>+...\simeq \cr
&\simeq &
-\frac{1}{4\pi\yb}\eta^{\alh\bt}f^{[\ul{ef}]}_{\bt\h\gm}\eta^{\h\gm\dl}N_{\ul{ef}}(0)
-\frac{1}{4\pi y}\eta^{\alh\bt}f^{[\ul{ef}]}_{\bt\h\gm}\eta^{\h\gm\dl}\Nhat_{\ul{ef}}(0).
\eea
%
%
% Jbar_3 Jbar_1
%
\bea
\JBalh(y)\JBdl(0)&\simeq &(\emph{cl}.) + \al^2
<\pbar\Xalh(y)\pbar\Xdl(0)>+... \simeq \cr &\simeq &
-\frac{1}{2\pi\yb^2}\eta^{\alh\dl}+
\frac{1}{4\pi}\frac{y}{\yb^2}\eta^{\alh\bt}f^{[\ul{ef}]}_{\bt\h\gm}\eta^{\h\gm\dl}N_{\ul{ef}}(0)-\frac{1}{4\pi\yb}\eta^{\alh\bt}f^{[\ul{ef}]}_{\bt\h\gm}\eta^{\h\gm\dl}
\Nhat_{\ul{ef}}(0).\cr &&
\eea
%
%
% COMMENTS ON THE OPE´S
%
All the current OPEs respect the $Z_4$-grading of the $psu(2,2|4)$
super-algebra. In fact the OPE of two currents with indices $A$
and $B$ is proportional to a current with index $C=A+B\pmod 4$. Of
course this reflects the fact that the tree-level interactions
between the $X$ fields respect the $Z_4$-automorphism of the
super-algebra, since the couplings which we can obtain are allowed
by the matrices (\ref{Amatrix}) and (\ref{Vmatrix}).

Using the above results we have checked
that the OPEs between the currents and the classical
equations of motion\footnote{The equations of motion are listed in
the appendix C.} derived from (\ref{initialaction}) vanish.\\
We have checked also that the OPEs found here reproduce commutators
of the currents computed in~\cite{bianchi} after the Wick rotation
to the Minkowskian world-sheet.  Because of our gauge choice $J_0$
and $\Jbar_0$ are absent in the commutators and the constraints,
which are present in ~\cite{bianchi}, vanish.

\section*{Acknowledgments}
I would like to thank Konstantin Zarembo for suggesting the
problem and for the guidance. I am also grateful to Gianluca
Grignani and Brenno Carlini Vallilo for helpful discussions. I
would also like to thank VR for partial financial support under
grant 621-2004-3178.

\appendix

\section*{Appendix A: Notation}
The $psu(2,2|4)$ super Lie algebra has a special inner symmetry, the so-called $\textbf{Z}_4$-auto\-morph\-ism~\cite{B9907200}, that allows to decompose it in
\be \label{algebradec}
\mathcal{G}=\mathcal{H}_0 \oplus\mathcal{H}_1\oplus\mathcal{H}_2\oplus\mathcal{H}_3.
\ee
The space $\mathcal{H}_k$ is the eigenspace with respect to the $Z_4$ action and the corresponding eigenvalue is $\imath^k$. Thus $\mathcal{H}_0$ is the locus of fixed points with respect the $Z_4$ transformation. Since the $Z_4$-grading is an automorphism of the super Lie algebra, the decomposition (\ref{algebradec}) respects the structure of the algebra, i.e. satisfies $ [\mathcal{H}_m,\mathcal{H}_n] \subset \mathcal{H}_{m+n\pmod 4}$ and also the bilinear form is $Z_4$-invariant:
\be
<\mathcal{H}_m,\mathcal{H}_n>=0 \qquad unless \;\; m+n=0\pmod 4
\ee
The subalgebra $\mathcal{H}_0$ is exactly the invariant subalgebra for the gauge $SO(4,1)\times SO(5)$ group. The $\mathcal{H}_2$ subalgebra is the space for the remaining bosonic elements (it contains the ``translation" generators), while $\mathcal{H}_1$ and $\mathcal{H}_3$ contain the fermionic elements (``supersymmetry" generators).\\
Therefore the generators of the super-algebra are decomposed in:
\be T_{\ula} \in \mathcal{H}_2 \qquad T_\al \in \mathcal{H}_1
\qquad T_{\alh} \in \mathcal{H}_3 \qquad T_{[\ulcd]} \in
\mathcal{H}_0, \ee and consequently the currents: \bea
J&=&g^{-1}\p g= J^A T_A= \Ja T_{\ula}+ \Jal T_\al + \Jalh T_{\alh}
+\Jzero T_{[\ulcd]}\cr \Jbar&=&g^{-1}\pbar g =\Jbar^A T_A=  \JBa
T_{\ula}+ \JBal T_\al + \JBalh T_{\alh} +\Jbzero T_{[\ulcd]}. \eea
The indices $A=(\ula, [\ulcd], \al,\alh)$ label the tangent spaces of the super Lie algebra; in particular $\ula= (a, a')$, $a=0,...,4$ labels the $so(4,1)$ vector index for $AdS_5$, $a'=5,...,9$ labels the $so(5)$ vector index for $S^5$, $[\ulcd]=([cd], [c'd'])$ and $\al$, $\alh=1,..16$ label the two sixteen-component Majorana-Weyl spinors in $D=10$. \\
In the curved background the two fermionic indices can couple
thanks to the matrix $\dl_{\al\alh}=(\gm^{01234})_{\al\alh}$, with $0,1,2,3,4$ the directions of $AdS_5$.
%
%
%\subsection{Structure Constants}
%
%

The supertrace is cyclic up
to a minus sign, i.e.
\be Str(XY)=(-1)^{deg(X)deg(Y)}Str(YX),
\ee
where $deg(X)=0$ if X is even and $deg(X)=1$ if X is odd. This is
consistent with the statistic.
The relation\footnote{The commutator has to be understood as a graded commutator: $[T_A, T_B]= T_AT_B-(-1)^{|A||B|}T_B T_A$, where $|A|=1$ for odd generators and $|A|=0$ for even generators.}
\be
Str(T_A[T_B,T_C])=Str([T_A,T_B]T_C),
\ee
furnish some important graded properties for the structure constants.
%
% List of the structure constants
%
The non-vanishing structure constants for the $psu(2,2|4)$ super
algebra are the following:
\bea
f^{[ab]}_{\al\bth}=\fraz(\gm^{[ab]})_{\al}^{\;\;\gm}\dl_{\gm\bth}\qquad
f^{[a'b']}_{\al\bth}=-\fraz(\gm^{[a'b']})_{\al}^{\;\;\gm}\dl_{\gm\bth}\qquad
f^{\al}_{[\ulcd]\bt}=-f^{\al}_{\bt[\ulcd]}=\fraz(\gmcd)_{\bt}^{\;\;\al}\quad\cr
f^{\alh}_{[\ulcd]\bth}=-f^{\alh}_{\bth[\ulcd]}=\fraz(\gmcd)_{\bth}^{\;\;\alh}\qquad
f^{\ula}_{\al\bt}=f^{\ula}_{\bt\al}=\gm^{\ula}_{\al\bt}\qquad
f^{\bth}_{\ula\bt}=-f^{\bth}_{\bt\ula}=-(\gma)_{\bt\gm}\dl^{\gm\bth}\cr
f^{\ula}_{\alh\bth}=f^{\ula}_{\alh\bth}=\gm^{\ula}_{\alh\bth}\qquad
f^{\al}_{\ula\alh}=-f^{\al}_{\alh\ula}=(\gma)_{\alh\bth}\dl^{\al\bth}\qquad
f^{[ef]}_{a b}=-f^{[ef]}_{b a}=\dl^{[e}_{a}\dl^{f]}_{b}\cr
f^{[e'f']}_{a' b'}=-f^{[e'f']}_{b'
a'}=-\dl^{[e'}_{a'}\dl^{f']}_{b'}\qquad
f^{\ul e}_{[\ulcd]\ul b}=-f^{\ul e}_{\ul b[\ulcd]}=\eta_{\ul b[\ul
c}\dl^{\ul e}_{\ul d]}\qquad\cr
f^{[\ul{gh}]}_{[\ul{cd}][\ul{ef}]}=\eta_{\ul c\ul e}\dl^{[\ul
g}_{\ul d}\dl^{\ul h]}_{\ul f}-\eta_{\ul c\ul f}\dl^{[\ul g}_{\ul
d}\dl^{\ul h]}_{\ul e}+\eta_{\ul d\ul f}\dl^{[\ul g}_{\ul
c}\dl^{\ul h]}_{\ul e}-\eta_{\ul d\ul e}\dl^{[\ul g}_{\ul
c}\dl^{\ul h]}_{\ul f}\qquad\qquad \eea
The metric $\eta_{A B}$ is given by:
\bea \eta_{\ulab}\qquad
\eta_{\al\bth}=-\eta_{\bth\al}=\dl_{\al\bth}\qquad
\eta_{[a'b'][c'd']}=-\eta_{a'[c'}\eta_{d']b'}\qquad
\eta_{[ab][cd]}=\eta_{a[c}\eta_{d]b} \eea
Furthermore if  a super-matrix is defined as:
\bea
K=
\left[\begin{array}{cc}
 A& C \\
D & B\\
\end{array}\right],
\eea
with $A$ and $B$ even matrices and $C$ and $D$ odd, then the supertranspose is given by:
\bea
K^{ST}=
\left[\begin{array}{cc}
 A^T& -D^T \\
C^T & B^T\\
\end{array}\right].
\eea
\subsection*{Gamma matrices in $D=10$ dimensions}
In $D=10$ dimensions in the reducible Majorana-Weyl representations the ($32\times 32$) Dirac gamma
matrices $\Gm^{\ul m}_{AB}$ are real and symmetric and they
consist of two symmetric $16\times 16$ matrices
$\gm_{\al\bt}^{\ul m}$, $\gm^{\ul m \;\al\bt}$ on the
off-diagonal\footnote{Since we are describing gamma matrices in a flat space we adopt the standard notation for the indices: $m=0,...,9$ is the $SO(9,1)$ vector index.}.
\bea \Gm^{\ul m} = \left[\begin{array}{cc}
0 & \gm^{\ul m \;\al\bt} \\
\gm_{\al\bt}^{\ul m}&0\\
\end{array}\right].
\eea
In the case of the type IIB superstring the two Majorana-Weyl
spinors have the same chirality, thus they transform in the same
$SO(9,1)$ representation.
Following~\cite{GSW} it is possible to construct explicitly the $\gm$ matrices from the
$SO(8)$ gamma matrices which themselves are direct product of
Pauli matrices:
\bea \gm^i_ {\al\bt}= \left[\begin{array}{cc}
0 & \sigma^{i\;a \dot{a}} \\
\sigma_{b \dot{b}}^{i}&0\\
\end{array}\right],
\eea
where $i=1,...,8$ and the $\sigma_{b \dot{b}}^{i}$ are the
antisymmetric real $SO(8)$ Pauli matrices and they satisfy the
following algebra:
\bea \sigma_{a \dot{a}}^{i}\sigma_{ \dot{a} b}^{j}+\sigma_{a
\dot{a}}^{j}\sigma_{\dot{a} b}^{i}=2\dl^{i j}\dl_{a b} \eea with
$\sigma_{\dot{a} b}^{i}$ the transpose of $\sigma_{ b
\dot{a}}^{i}$\footnote{A specific set for the $\sigma$ matrices is given in~\cite{GSW}:
\bea
&&\sigma^1=\epsilon\times \epsilon\times \epsilon \qquad
\sigma^2= 1\times \tau_1\times \epsilon \qquad
\sigma^3= 1\times \tau_3\times \epsilon \qquad
\sigma^4= \tau_1\times \epsilon\times 1 \cr
&&\sigma^5= \tau_3\times \epsilon\times 1\qquad
\sigma^6= \epsilon\times 1\times \tau_1\qquad
\sigma^7=\epsilon\times 1\times \tau_3\qquad
\sigma^8=1\times 1\times 1,
\eea
where $\tau_i$ are the Pauli matrices and $\epsilon=\imath\tau_2$.}.
A ninth one that anti-commutes with these eight is given
by~\cite{Chesterman}:
\bea \gm^{9}_{\al\bt}=\gm^{9\; \al\bt}= \left[\begin{array}{cc}
1_{8} & 0 \\
0 & -1_8\\
\end{array}\right],
\eea
and the values of $\gm^{0\,\al\bt}$ and $\gm^{0}_{\al\bt}$
are similarly defined in order to be consistent with their
algebra:
\bea \gm^{0}_{\al\bt}= \left[\begin{array}{cc}
-1_8 & 0 \\
0 & -1_8\\
\end{array}\right],\qquad
\gm^{0\; \al\bt}= \left[\begin{array}{cc}
1_8 & 0 \\
0 & 1_8\\
\end{array}\right].
\eea
\section*{Appendix B: OPE}

In this section the OPEs of the matter currents will be treated explicitly. The overall factor $\al^2$ is omitted in the final results.

%\subsection{$J_2 J_2$}

\bea
&&\Ja(y)\Jd(z)\simeq \al^2
 <\p \Xa (y)\p \Xd(z)>+...=\cr
&&=  -\frac{1}{2\pi}\eta^{\ul{ad}}\p_y \p_z \log|y-z|^2+\frac{1}{16\pi^2} \eta^{[\ul{ad}][\ul{ef}]}\int d^2 \om \ga -\p_y\pbar_{\omb}\log|y-\om|^2 N_{\ul{ef}}(\om)\p_z \log|\om-z|^2+\cr
&&+\p_y \log|y-\om|^2 N_{\ul{ef}}(\om)\p_z\pbar_{\omb} \log|\om-z|^2\gc+\cr
&&+\frac{1}{16\pi^2}\eta^{[\ul{ad}][\ul{ef}]}\int d^2\om \ga -\p_y\p_\om \log|y-\om|^2\Nhat_{\ul{ef}}(\om)\p_z \log|\om-z|^2+\cr
&&+\p_y \log|y-\om|^2\Nhat_{\ul{ef}}(\om)\p_z\p_\om \log|\om-z|^2\gc=
\eea
The integrand containing $N_{\ul{ef}}$ is a $\dl$ function (up to a minus sign) and so it can be easily integrated. Furthermore all the currents are expanded around $z$, i.e. $\Nhat_{\ul{ef}}(\om)\cong \Nhat_{\ul{ef}}(z)+...$ and $N_{\ul{ef}}(\om)\cong N_{\ul{ef}}(z)+...$, the terms with the derivatives of the currents can be neglected at this
order, just by dimensional analysis. Setting $z=0$ the OPE becomes:
\bea
&&\Ja(y)\Jd(0)\simeq
 -\frac{\eta^{\ul{ad}}}{2\pi}\frac{1}{y^2}-\frac{1}{4\pi y}\eta^{[\ul{ad}][\ul{ef}]}N_{\ul{ef}}(0)+\frac{1}{4\pi}\eta^{[\ul{ad}][\ul{ef}]}\Nhat_{\ul{ef}}(0)\frac{\yb}{y^2}.
\eea
%
% J_2 Jbar_2
%
\bea
&&\Ja(y)\JBd(z)\simeq (cl) +\al^2 <\p \Xa (y)\pbar
\Xd(z)>+...=\cr
&&= -\frac{1}{2\pi}\eta^{\ul{ad}}\p_y\pbar_{\zb}\log|y-z|^2+\frac{1}{16\pi^2}\eta^{[\ul{ad}][\ul{ef}]}\int d^2 \om \ga
-\p_y\pbar_{\omb}\log|y-\om|^2 N_{\ul{ef}}(\om)\pbar_{\zb}\log|\om-z|^2+\cr
&&+\p_y \log|y-\om|^2 N_{\ul{ef}}(\om)\pbar_{\zb}\pbar_{\omb}
\log|\om-z|^2\gc+\cr
&&+\frac{1}{16\pi^2}\eta^{[\ul{ad}][\ul{ef}]}\int d^2\om\ga -\p_y\p_\om
\log|y-\om|^2\Nhat_{\ul{ef}}(\om)\pbar_{\zb}\log|\om-z|^2+\cr
&&+\p_y \log|y-\om|^2\Nhat_{\ul{ef}}(\om)\pbar_{\zb}\p_\om \log|\om-z|^2\gc =\cr
&&\simeq -\frac{1}{4\pi (\yb-\zb)}\eta^{[\ul{ad}][\ul{ef}]}N_{\ul{ef}}(z)-\frac{1}{4\pi (y-z)}\eta^{[\ul{ad}][\ul{ef}]}\Nhat_{\ul{ef}}(z),
\eea
where the first term in the second line is a $\dl$ function and it will be not considered here, since only singular terms are taken in account; the result (\ref{j2jbar2}) is obtained with $z=0$.
Since the procedure is completely analogous for the remaining bosonic components of the currents, we will not rewrite them, the results are listed in (\ref{jbar2j2}), (\ref{jbar2jbar2}).
%
%
% Jbar_2 J_2
%
%\bea
%&&\JBa(y)\Jd(0)\;\cong (cl) +\al^2 ( <\pbar \Xa (y)\p
%\Xd(0)>+...)=\cr
%
%&&= -\frac{1}{4\pi\yb}\eta^{[\ul{ad}][\ul{ef}]}N_{\ul{ef}}(z)-\frac{1}{4\pi %y}\eta^{[\ul{ad}][\ul{ef}]}\Nhat_{\ul{ef}}(0).
%\eea
%
%
% Jbar_2 Jbar_2
%
%\bea
%&&\JBa(y)\JBd(0)\;\cong (cl) +\al^2 ( <\pbar \Xa (y)\pbar
%\Xd(0)>+...)=\cr
%
%&&=-\frac{1}{2\pi\yb^2}\eta^{\ul{ad}}+\frac{1}{4\pi}\frac{y}{\yb^2}\eta^{[\ul{ad}][\ul{ef}]}
%N_{\ul{ef}}(0) -\frac{1}{4\pi \yb}\eta^{[\ul{ad}][\ul{ef}]}\Nhat_{\ul{ef}}(0)
%\eea
%
%\subsection{$J_2 J_1$}
%

In the case of the OPE between $J_2 J_1$, $J_2 J_3$, $J_1 J_1$ and
$J_3 J_3$ there are contributions from the commutators in
(\ref{opej}), since the $X$ fields can propagate "freely", as one
can understand from the entries of the matrix (\ref{Amatrix}) and
from the underlying super-algebra. We present explicitly  only the
OPE for the $J J$ components in each case, since for the other
components the OPEs are completely analogous.
% and the results for
%them are in (\ref{j2jbar1}), (\ref{jbar2j1}), (\ref{jbar2jbar1}).
%
\bea
\Ja(y)\Jdl(0)\simeq \al^2 (<\p\Xa(y)\p\Xdl(0)>+
<\p\Xa(y)[\Jcl_3,X_2]^\dl (0)>+
<[\Jcl_3,X_3]^{\ula}(y)\p\Xdl(0)>+...) \eea
The first term is
\bea
&&<\p\Xa(y)\p\Xdl(0)>=\p_y\p_z <\Xa (y)\Xdl(z)>|_{z=0}=\cr
&&=\frac{1}{8\pi^2} f^{\ula}_{\h\gm\h\rho}\eta^{\h\gm\dl}\int d^2 \om \ga
-\p_y\p_\om \log|y-\om|^2\Jbar^{\h\rho}(\om)\p_z\log|\om-z|^2+\p_y \log|y-\om|^2\Jbar^{\h\rho}\p_z\p_\om \log|\om-z|^2\gc |_{z=0}=\cr &&=
\frac{1}{2\pi}\frac{\yb}{y^2}f^{\ula}_{\h\gm\h\rho}\eta^{\h\gm\dl}\Jbar^{\h\rho}(0),
\eea
the second term:
\bea
\p\Xa(y)[J_3,X_2]^\dl(0)&=&
-\p\Xa(y)f^\dl_{\ulb\h\rho}\Xb(0)J^{\h\rho}(0)=
\frac{1}{2\pi}\eta^{\ulab}\p_y\log|y-z|^2 f^{\rho}_{\ulb\h\rho}J^{\h\rho}(z)|_{z=0}=\cr
&=&\frac{1}{2\pi y} f^{\ula}_{\h\rho\h\gm}\eta^{\h\gm\dl}J^{\h\rho}(0),
\eea
and the last term is:
\bea [J_3,X_3]^{\ula}(y)\p\Xdl(0)&=&
f^{\ula}_{\h\rho\h\gm}J^{\h\rho}(y)X^{\h\gm}(y)\p\Xdl(0)=
-\frac{1}{2\pi}
f^{\ula}_{\h\rho\h\gm}J^{\h\rho}(y)\eta^{\h\gm\dl}\p_z
\log|y-z|^2|_{z=0}=\cr &=& \frac{1}{2\pi y}
f^{\ula}_{\h\rho\h\gm}J^{\h\rho}(0)\eta^{\h\gm\dl}. \eea
Therefore one gets:
\bea
&&\Ja(y)\Jdl(0) \simeq
\frac{1}{2\pi}\frac{\yb}{y^2}f^{\ula}_{\h\gm\h\rho}\eta^{\h\gm\dl}\Jbar^{\h\rho}(0)+
\frac{1}{\pi y}
f^{\ula}_{\h\rho\h\gm}\eta^{\h\gm\dl}J^{\h\rho}(0).
\eea
%
%
% J_2 Jbar_1
%
%\bea
%\Ja(y)\JBdl(0) \;&\cong &
%\al^2 (<\p\Xa(y)\pbar\Xdl(0)>+
%<\p\Xa(y)[\Jbarcl_3,X_2]^\dl (0)+
%[\Jcl_3,X_3]^{\ula}(y)\pbar\Xdl(0)+...)\cr &&
%\cong -\frac{1}{2\pi y}f^{\ula}_{\h\gm\h\rho}\eta^{\h\gm\dl}\Jbar^{\h\rho}(0)+ \frac{1}{2\pi %y}f^{\ula}_{\h\gm\h\rho}\eta^{\h\gm\dl}\Jbar^{\h\rho}(0)+ \frac{1}{2\pi \yb}\eta^{\h\gm\dl}J^{\h\rho}(0)=
%\frac{1}{2\pi \yb}\eta^{\h\gm\dl}J^{\h\rho}(0) .\cr
%&&
%\eea
%
% Jbar_2 J_1
%
%\bea
%&&\JBa(y)\Jdl(0)\; \cong
%\al^2 (<\pbar\Xa(y)\p\Xdl(0)>+<\pbar\Xa(y)[\Jcl_3,X_2]^\dl (0)+[\Jbarcl_3,X_3]^{\ula}(y)\p\Xdl(0)+...)=\cr
%
%&&=-\frac{1}{2\pi y} f^{\ula}_{\h\gm\h\rho}\eta^{\h\gm\dl}\Jbar^{\h\rho}(0)+
%
%\frac{1}{2\pi \yb}f^{\ula}_{\h\gm\h\rho}\eta^{\h\gm\dl}J^{\h\rho}(0)
%%
%+\frac{1}{2\pi y}f^{\ula}_{\h\gm\h\rho}\eta^{\h\gm\dl}\Jbar^{\h\rho}(0)=
%
%\frac{1}{2\pi \yb}f^{\ula}_{\h\gm\h\rho}\eta^{\h\gm\dl}J^{\h\rho}(0)
%.\eea
%
%
% Jbar_2 Jbar_1
%
%\bea
%&&\JBa(y)\JBdl(0)\; \cong  \al^2
%(<\pbar\Xa(y)\pbar\Xdl(0)>+ <\pbar\Xa(y)[\Jbar_3,X_2]^\dl (0)+[\Jbar_3,X_3]^{\ula}(y)\pbar\Xdl(0)+...)=\cr
%
%
%&&=
%\frac{1}{2\pi\yb}f^{\ula}_{\h\gm\h\rho}\eta^{\h\gm\dl}\Jbar^{\h\rho}(0)
%+\frac{1}{2\pi\yb}f^{\ula}_{\h\gm\h\rho}\eta^{\h\gm\dl}\Jbar^{\h\rho}(0)
%-\frac{1}{2\pi\yb}f^{\ula}_{\h\gm\h\rho}\eta^{\h\gm\dl}\Jbar^{\h\rho}(0)=
%\frac{1}{2\pi\yb}f^{\ula}_{\h\gm\h\rho}\eta^{\h\gm\dl}\Jbar^{\h\rho}(0).\cr
%&&
%\eea
%
%
%

%\subsection{$J_2 J_3$}
%
% J_2 J_3
%
\bea
\Ja(y)\Jdlh (0)\simeq \al^2( <\p\Xa(y)\p\Xdlh
(0)>+\p\Xa(y)[\Jcl_1,X_2]^{\h\dl}(0)+[\Jcl_1,X_1]^{\ula}(y)\p\Xdlh
(0)+...). \cr
&&
\eea
The first term:
\bea \label{dxadxdeltah}
&&<\p\Xa(y)\p\Xdlh(0)>=-\p_y\p_z(A^{-1}VA^{-1})^{\ula\h\dl}|_{z=0}=\cr
&&=\frac{1}{8\pi^2}f^{\ula}_{\rho\gm}\eta^{\gm\h\dl}\int d^2
\om\ga -\p_y\pbar_{\omb}\log|y-\om|^2 J^\rho
(\om)\p_z\log|\om-z|^2+\p_y
\log|y-\om|^2J^\rho(\om)\pbar_{\omb}\p_z\log|\om-z|^2\gc|_{z=0}\cr &&
\eea
The second term: \bea
-\p\Xa(y)f^{\h\dl}_{\ulb\rho}\Xb(0)J^\rho(0)=\frac{1}{2\pi}\eta^{\ulab}f^{\h\dl}_{\ulb\rho}J^\rho(z)\p_y
\log|y-z|^2|_{z=0}=\frac{1}{2\pi}f^{\ula}_{\rho\gm}\eta^{\gm\h\dl}J^\rho(0)\frac{1}{y}
\eea
The third term:
\bea
f^{\ula}_{\rho\gm}J^\rho(y)\Xdl(y)\p\Xdlh(0)=-\frac{1}{2\pi}
f^{\ula}_{\rho\gm}\eta^{\gm\h\dl}J^\rho(y)\p_z\log|y-z|^2|_{z=0}
\eea
Thus the OPE is:
\bea
&&\Ja(y)\Jdlh (0)\simeq -\frac{1}{2\pi
y}f^{\ula}_{\rho\gm}\eta^{\gm\h\dl}J^\rho(0)+ \frac{1}{2\pi
y}f^{\ula}_{\rho\gm}\eta^{\gm\h\dl}J^\rho(0)+ \frac{1}{2\pi
y}f^{\ula}_{\rho\gm}\eta^{\gm\h\dl}J^\rho(0)\simeq \frac{1}{2\pi
y}f^{\ula}_{\rho\gm}\eta^{\gm\h\dl}J^\rho(0).\cr && \eea
%
%
%% J_2 Jbar_3
%
%\bea
%&&\Ja(y)\JBdlh(0)\cong
%\al^2 (<\p\Xa(y)\pbar\Xdlh(0)> +
%\p\Xa(y)[\Jbar_1,X_2]^{\h\dl}(0)+[J_1,X_1]^{\ula}(y)\pbar\Xdlh(0)+...)=\cr
%
%&&= -\frac{1}{2\pi \yb} f^{\ula}_{\rho\gm}\eta^{\gm\h\dl}J^\rho (0)+
%\frac{1}{2\pi y} f^{\ula}_{\rho\gm}\eta^{\gm\h\dl}\Jbar^\rho(0)+
%\frac{1}{2\pi\yb}f^{\ula}_{\rho\gm}J^\rho(0)\eta^{\gm\h\dl}=\cr
%&&= \frac{1}{2\pi y} f^{\ula}_{\rho\gm}\eta^{\gm\h\dl}\Jbar^\rho(0).
%\eea
%
% Jbar_2 J_3
%
%\bea
%&&\JBa(y)\Jdlh(0)\cong \al^2 (<\pbar\Xa(y)\p\Xdlh(0)> +
%\pbar\Xa(y)[J_1,X_2]^{\h\dl}(0)+[\Jbar_1,X_1]^{\ula}(y)\p\Xdlh(0)+...)=\cr
%
%&&=-\frac{1}{2\pi\yb}f^{\ula}_{\rho\gm}\eta^{\gm\h\dl}J^\rho(0)+
%\frac{1}{2\pi\yb}f^{\ula}_{\rho\gm}\eta^{\gm\h\dl}J^\rho(0)+
%\frac{1}{2\pi y}f^{\ula}_{\rho\gm}\eta^{\gm\h\dl}\Jbar^\rho(0)=
%\frac{1}{2\pi y}f^{\ula}_{\rho\gm}\eta^{\gm\h\dl}\Jbar^\rho(0).
%\eea
%
%
% Jbar_2 Jbar_3
%
%\bea
%&&\JBa(y)\JBdlh(0)\; \cong
%\al^2 (<\pbar\Xa(y)\pbar\Xdlh(0)>+\pbar\Xa(y)[\Jbar_1,X_2]^{\h\dl}(0)+[\Jbar_1,X_1]^{\ula}(y)\pbar\Xdlh(0)+...)=\cr
%
%&&= \frac{1}{2\pi}\frac{y}{\yb^2}f^{\ula}_{\rho\gm}\eta^{\gm\h\dl}J^\rho(0)+ \frac{1}{\pi\yb} f^{\ula}_{\rho\gm}\Jbar^\rho (0)\eta^{\gm\h\dl}.
%\eea
%
%
%
%

%\subsection{$J_1 J_1$}
%
% J_1 J_1
%
\bea
\Jal(y)\Jdl(0)\simeq
\al^2(<\p\Xal(y)\p\Xdl(0)>+\p\Xal(y)[J_2,X_3]^\dl(0)+[J_2,X_3]^\al(y)\p\Xdl(0)+...)\cr
&&
\eea
The first term:
\bea
&&<\p\Xal(y)\p\Xdl(0)>=-\p_y\p_z(A^{-1}VA^{-1})^{\al\dl}|_{z=0}=\cr
&&=\frac{1}{8\pi^2}f^{\al}_{\ul l\h\gm}\eta^{\h\gm\dl}\int
d^2\om\ga -\p_y\p_\om \log|y-\om|^2\Jbar^{\ul l}(\om)\p_z
\log|\om-z|^2+\p_y \log|y-\om|^2\Jbar^{\ul
l}(\om)\p_z\p_\om  \log|\om-z|^2\gc|_{z=0} =\cr &&=
\frac{1}{2\pi}\frac{\yb}{y^2}f^{\al}_{\ul
l\h\gm}\eta^{\h\gm\dl}\Jbar^{\ul l}(0) \eea
The second term:
\bea
\p\Xal(y)f^{\dl}_{\ul l \bth}J^{\ul
l}(0)\Xbth(0)=-\frac{1}{2\pi}\eta^{\al\bth}f^{\dl}_{\ul l \bth}J^{\ul l}(z)\p_y
\log|y-z|^2|_{z=0}
=\frac{1}{2\pi} f^{\al}_{\ul l\h\gm}\eta^{\h\gm\dl}J^{\ul l}(0)\frac{1}{y}
\eea
The third term:
\bea
f^{\al}_{\ul l \bth}J^{\ul l}(y)\Xbth(y)\p\Xdl(0)=-\frac{1}{2\pi}
f^{\al}_{\ul l \bth}J^{\ul l}(y)\eta^{\bth\dl}\p_z \log|y-z|^2|_{z=0}
\eea
Therefore the OPE is given by:
\bea
&&\Jal(y)\Jdl(0)\simeq
\frac{1}{2\pi}\frac{\yb}{y^2}f^{\al}_{\ul l\h\gm}\eta^{\h\gm\dl}\Jbar^{\ul l}(0)+
\frac{1}{\pi y} f^{\al}_{\ul l\h\gm}\eta^{\h\gm\dl}J^{\ul l}(0).
\eea
%
%
% Jbar_1 J_1
%
%\bea
%&&\JBal(y)\Jdl(0)\; \cong
%\al^2\ga <\pbar\Xal(y)\p\Xdl(0)>+\pbar\Xal(y)[J-2,X_3]^\dl(0)+[\Jbar_2,X_3]^\al(y)\p\Xdl(0)+...\gc=
%\cr
%&&=
%-\frac{1}{2\pi y}f^{\al}_{\ul l\h\gm}\eta^{\h\gm\dl}\Jbar^{\ul l}(0)+
%\frac{1}{2\pi\yb} f^{\al}_{\ul l\h\gm}\eta^{\h\gm\dl}J^{\ul l}(0)+
%\frac{1}{2\pi y}f^{\al}_{\ul l\bth}\Jbar^{\ul l}(y) \eta^{\bth \dl}= \frac{1}{2\pi\yb} f^{\al}_{\ul l\h\gm}\eta^{\h\gm\dl}J^{\ul l}(0).
%\eea
%
%
% J_1 Jbar_1
%
%\bea
%&&\Jal(y)\JBdl(0) \;\cong
%\al^2(<\p\Xal(y)\pbar\Xdl(0)>+\p\Xal(y)[\Jbar_2,X_3]^\dl(0)+[J_2,X_3]^\al(y)\pbar\Xdl(0)+...)=\cr
%
%&&=-\frac{1}{2\pi y}f^{\al}_{\ul l\h\gm}\eta^{\h\gm\dl}\Jbar^{\ul l}(0)+
%\frac{1}{2\pi y} f^{\al}_{\ul l\h\gm}\eta^{\h\gm\dl}\Jbar^{\ul l}(0)+
%\frac{1}{2\pi \yb} f^{\al}_{\ul l\bth}J^{\ul l}(0)\eta^{\bth \dl}=
%\frac{1}{2\pi \yb} f^{\al}_{\ul l\bth}J^{\ul l}(0)\eta^{\bth \dl}.
%\eea
%
%
% Jbar_1 Jbar_1
%
%\bea
%&&\JBal(y)\JBdl(0)\; \cong
%\al^2(<\pbar\Xal(y)\pbar\Xdl(0)>+\pbar\Xal(y)[\Jbar_2,X_3]^\dl(0)+[\Jbar_2,X_3]^\al(y)\pbar\Xdl(0)+...)=\cr
%
%&&= \frac{1}{2\pi\yb}f^{\al}_{\ul l\h\gm}\eta^{\h\gm\dl}\Jbar^{\ul l}(0)+
%\frac{1}{2\pi\yb}f^{\al}_{\ul l\bth}\Jbar^{\ul l}(0)\eta^{\bth \dl}
%-\frac{1}{2\pi\yb}f^{\al}_{\ul l\bth}\eta^{\bth \dl}\Jbar^{\ul l}(0)=
%\frac{1}{2\pi\yb}f^{\al}_{\ul l\bth}\eta^{\bth \dl}\Jbar^{\ul l}(0).
%\eea
%
%
%\subsection{$J_3 J_3$}
%
% J_3 J_3
%
\bea
\Jalh(y)\Jdlh(0)\simeq \al^2 \ga
<\p\Xalh(y)\p\Xdlh(0)>+\p\Xalh(y)[J_2,X_1]^{\h\dl}(0)+[J_2,X_1]^{\alh}(y)\p\Xdlh(0)+...\gc\cr &&
\eea
The first term is:
\bea
&&<\p\Xalh(y)\p\Xdlh(0)>=-\p_y\p_z(A^{-1}VA^{-1})^{\alh\h\dl}|_{z=0}=\cr
&&=\frac{1}{8\pi^2}f^{\alh}_{\ul l\gm}\eta^{\gm\h\dl}\int d^2\om \ga
-\p_y\pbar_{\omb}\log|y-\om|^2 J^{\ul l}(\om)\p_z \log|\om-z|^2 + \p_y \log|y-\om|^2 J^{\ul
l}(\om)\p_z\pbar_{\omb}\log|\om-z|^2 \gc|_{z=0}\cr &&
\eea
The second term is:
\bea
\p\Xalh(y)f^{\h\dl}_{\ul l\bt}J^{\ul
l}(0)\Xbt(0)=\frac{1}{2\pi}\eta^{\alh\bt}f^{\h\dl}_{\bt\ul l}J^{\ul l}(z)\p_y
\log|y-z|^2|_{z=0}=
\frac{1}{2\pi y} f^{\alh}_{\ul l\gm}\eta^{\gm\h\dl}J^{\ul l}(0)
\eea
The third term:
\bea [J_2,X_1]^{\alh}(y)\p\Xdlh(0)=-\frac{1}{2\pi}f^{\alh}_{\ul l\gm}\eta^{\gm\h\dl}J^{\ul
l}(y)\p_z\log|y-z|^2|_{z=0}\eea
Thus the OPE is:
\bea
&& \Jalh(y)\Jdlh(0)\simeq
-\frac{1}{2\pi y} f^{\alh}_{\ul l\gm}\eta^{\gm\h\dl}J^{\ul l}(0)
\frac{1}{2\pi y} f^{\alh}_{\ul l\gm}\eta^{\gm\h\dl}J^{\ul l}(0)+
\frac{1}{2\pi y}f^{\alh}_{\ul l\gm}\eta^{\gm\h\dl}J^{\ul l}(0)\cong \cr
&&\cong
\frac{1}{2\pi y} f^{\alh}_{\ul l\gm}\eta^{\gm\h\dl}J^{\ul l}(0). \eea
%
%
%\bea
%&& \JBalh(y)\Jdlh(0) \cong \al^2 \ga
%<\pbar\Xalh(y)\p\Xdlh(0)>+\pbar\Xalh(y)[J_2,X_1]^{\h\dl}(0)+[\Jbar_2,X_1]^{\alh}(y)\p\Xdlh(0)+...\gc=\cr
%&&
%=
%-\frac{1}{2\pi\yb}f^{\alh}_{\ul l\gm}\eta^{\gm\h\dl}J^{\ul l}(0) +
%\frac{1}{2\pi\yb} f^{\alh}_{\ul l\gm}\eta^{\gm\h\dl}J^{\ul l}(0)+
%\frac{1}{2\pi y}f^{\alh}_{\ul l\gm}\Jbar^{\ul l}(0)\eta^{\gm\h\dl}=
%\frac{1}{2\pi y}f^{\alh}_{\ul l\gm}\Jbar^{\ul l}(0)\eta^{\gm\h\dl}.
%\eea
%
% J_3 Jbar_3
%
%\bea
%&& \Jalh(y)\JBdlh(0)\; \cong \al^2 \ga
%<\p\Xalh(y)\pbar\Xdlh(0)> +
%\p\Xalh(y)[\Jbar_2,X_1]^{\h\dl}(0)+[J_2,X_1]^{\alh}(y)\pbar\Xdlh(0)+...\gc
%= \cr &&
%
%=-\frac{1}{2\pi\yb}f^{\alh}_{\ul l\gm}\eta^{\gm\h\dl}J^{\ul l}(0)+
%\frac{1}{2\pi y} f^{\alh}_{\ul l\gm}\eta^{\gm\h\dl}\Jbar^{\ul l}(0)+
%\frac{1}{2\pi\yb}f^{\alh}_{\ul l\gm}J^{\ul l}(0)\eta^{\gm\h\dl}=
%\frac{1}{2\pi y} f^{\alh}_{\ul l\gm}\eta^{\gm\h\dl}\Jbar^{\ul l}(0).
%\eea
%
% Jbar_3 Jbar_3
%
%\bea
%&& \JBalh(y)\JBdlh(0) \;\cong \al^2 \ga
%<\pbar\Xalh(y)\pbar\Xdlh(0)> +
%\pbar\Xalh(y)[\Jbar_2,X_1]^{\h\dl}(0)+[\Jbar_2,X_1]^{\alh}(y)\pbar\Xdlh(0)+...\gc
%= \cr &&
%
%= \frac{1}{2\pi}\frac{y}{\yb^2}f^{\alh}_{\ul l\gm}\eta^{\gm\h\dl}J^{\ul l}(0)+
%\frac{1}{2\pi\yb} f^{\alh}_{\ul l\gm}\eta^{\gm\h\dl}\Jbar^{\ul l}(0)+
%\frac{1}{2\pi\yb} f^{\alh}_{\ul l\gm}\Jbar^{\ul l}(0)\eta^{\gm\h\dl}= \cr
%&&=\frac{1}{2\pi}\frac{y}{\yb^2}f^{\alh}_{\ul l\gm}\eta^{\gm\h\dl}J^{\ul l}(0)+
%\frac{1}{\pi\yb} f^{\alh}_{\ul l\gm}\eta^{\gm\h\dl}\Jbar^{\ul l}(0).
%\eea
%
%
%
%
%\subsection{$J_3 J_1$}
%
In the case of $J_3 J_1$ OPE the same algebraic structure of $J_2
J_2$ is involved, therefore we present briefly the result.
% J_3 J_1
%
%
\bea
&&\Jalh(y)\Jdl(0)\simeq (\textsl{cl}.) + \al^2
<\p\Xalh(y)\p\Xdl(0)>+... \simeq\cr
&&\simeq
\big(-\frac{1}{2\pi} \eta^{\alh\dl}\p_z\p_y
\log|y-z|^2+\frac{1}{16\pi^2}\eta^{\alh\bt}f^{[\ul{ef}]}_{\bt\h\gm}\eta^{\h\gm\dl}\int
d^2\om [ -\p_y\pbar_{\omb}\log|y-\om|^2 N_{\ul{ef}}(\om)\p_z
\log|\om-z|^2+\cr &&+\p_y\log|y-\om|^2
N_{\ul{ef}}(\om)\p_z\pbar_{\omb}\log|\om-z|^2-\p_y\p_\om
\log|y-\om|^2\Nhat_{\ul{ef}}(\om)\p_z \log|\om-z|^2+\cr &&+\p_y
\log|y-\om|^2\Nhat_{\ul{ef}}(\om)\p_z\p_\om \log|\om-z|^2]\big)|_{z=0}\simeq \cr
&&\simeq
-\frac{1}{2\pi y^2}\eta^{\alh\dl}+
\frac{1}{4\pi}\frac{\yb}{y^2}\eta^{\alh\bt}f^{[\ul{ef}]}_{\bt\h\gm}\eta^{\h\gm\dl}
\Nhat_{\ul{ef}}(0)
-\frac{1}{4\pi y}\eta^{\alh\bt}f^{[\ul{ef}]}_{\bt\h\gm}\eta^{\h\gm\dl}N_{\ul{ef}}(0),
\eea
where the classical term is omitted.

%
% Jbar_3 J_1
%
%\bea
%&&\JBalh(y)\Jdl(0)\; \cong (\emph{cl.}) + \al^2 \ga <\pbar\Xalh(y)\p\Xdl(z)>+...\gc = \cr
%&&
%=-\frac{1}{4\pi \yb}\eta^{\alh\bt}f^{[\ul{ef}]}_{\bt\h\gm}\eta^{\h\gm\dl}  N_{\ul{ef}}(0)-\frac{1}{4\pi y}\eta^{\alh\bt}f^{[\ul{ef}]}_{\bt\h\gm}\eta^{\h\gm\dl}\Nhat_{\ul{ef}}(0),
%\eea
%where this time also the local term has been omitted.
%
%
% J_3 Jbar_1
%
%\bea
%&&\Jalh(y)\JBdl(0)\;\cong
%(\emph{cl.}) + \al^2 \ga <\p\Xalh(y)\pbar\Xdl(z)>+...\gc = \cr
%&&=
%-\frac{1}{4\pi\yb}\eta^{\alh\bt}f^{[\ul{ef}]}_{\bt\h\gm}\eta^{\h\gm\dl}N_{\ul{ef}}(0)
%-\frac{1}{4\pi y}\eta^{\alh\bt}f^{[\ul{ef}]}_{\bt\h\gm}\eta^{\h\gm\dl}\Nhat_{\ul{ef}}(0).
%\eea
%
%
% Jbar_3 Jbar_1
%
%For the last components:
%\bea
%&&\JBalh(y)\JBdl(0)\;\cong
%(\emph{cl}.) + \al^2 \ga<\pbar\Xalh(y)\pbar\Xdl(0)>+...\gc =\cr
%&&=
%-\frac{1}{2\pi\yb^2}\eta^{\alh\dl}+ %\frac{1}{4\pi}\frac{y}{\yb^2}\eta^{\alh\bt}f^{[\ul{ef}]}_{\bt\h\gm}\eta^{\h\gm\dl}\Nhat_{\ul{ef}}(0)
%-\frac{1}{4\pi \yb}\eta^{\alh\bt}f^{[\ul{ef}]}_{\bt\h\gm}\eta^{\h\gm\dl}N_{\ul{ef}}(0).
%\eea
%
%
\section*{Appendix C: Classical Equations of Motion}
We now derive the classical equations of motion from the action
(\ref{initialaction}). Under small variations of the fields $g$ the currents satisfy~\cite{V0307018,B9907200}:
\bea \label{variation} \dl g=g X \qquad \dl g^{-1}=-Xg^{-1},
\qquad X \in H_i \cr \dl J_i=\p X +[J,X]_i \qquad \Jbar_i=\pbar X
+[\Jbar, X]_i \cr \dl J_0=[J,X]_0 \qquad \dl\Jbar_0=[\Jbar,X]_0
\cr \dl N=[N,\Lm]\qquad \dl\Nhat =[\Nhat,\h\Lm], \eea
where for the variation of the Lorentz ghost currents the gauge transformation is used since this is the most general
covariant variation which respects the $SO(4,1)\times SO(5)$
symmetry~\cite{V0307018}. Furthermore under gauge transformation
for the pure spinor actions $S_\lm$ and $\h S_{\h\lm}$
we have $\dl S_\lm=-N^{[\ulcd]}\p\Lm_{[\ulcd]}$ and $\dl\h
S_{\h\lm}=-\Nhat\pbar\h\Lm_{[\ulcd]}$. Plugging (\ref{variation})
in the action (\ref{initialaction}) and using the Maurer-Cartan
identities $\p\Jbar-\pbar J+[J,\Jbar]=0$ one gets:
\bea \label{classicaleom}
&& \nabar J_2=[J_3,\Jbar_3]+\fraz [N,\Jbar_2]-\fraz[J_2,\Nhat]\cr
&&\na \Jbar_2=-[J_1,\Jbar_1]+\fraz [N,\Jbar_2]-\fraz[J_2,\Nhat]
\cr && \nabar J_3=\fraz[N,\Jbar_3]-\fraz[J_3,\Nhat] \cr && \na
\Jbar_3=
-[J_1,\Jbar_2]-[J_2,\Jbar_1]+\fraz[N,\Jbar_3]-\fraz[J_3,\Nhat] \cr
&& \nabar
J_1=[J_3,\Jbar_2]+[J_2,\Jbar_3]+\fraz[N,\Jbar_1]-\fraz[J_1,\Nhat]\cr
&& \na \Jbar_1=\fraz[N,\Jbar_1]-\fraz[J_1,\Nhat]\cr && \nabar
N=\fraz [N,\Nhat] \cr && \na \Nhat=-\fraz [N,\Nhat], \eea
where the covariant derivatives are $\na=\p +[J_{0},\;\:]$ and $\nabar=\pbar+[\Jbar_{0},\;\:]$.

\end{document}